\documentclass[referee]{raa}            
\usepackage{xcolor}
\usepackage{graphicx,times}             
\usepackage{natbib}
\usepackage{amssymb,amsmath}
\bibpunct{(}{)}{;}{a}{}{,}

\usepackage[pagebackref=true]{hyperref}

\begin{document}

  \title{A multi-epoch spectral and photometric study to understand the nature of the interesting Be star candidate HD 249179
}

   \volnopage{Vol.0 (20xx) No.0, 000--000}      
   \setcounter{page}{1}          

   \author{Suman Bhattacharyya 
      \inst{1}
   \and Blesson Mathew
      \inst{1}
   \and S Muneer
      \inst{2}
   \and Sreeja S Kartha
      \inst{1}
   \and Prakruthi Prasad
      \inst{1}
   }

   \institute{CHRIST (Deemed to be University), Bangalore, India\\
        \and
             Indian Institute of Astrophysics (IIA), Bangalore, India\\
\vs\no
   {\small Received 20xx month day; accepted 20xx month day}}

\abstract{HD~249179, a B5 star with unclear classification as either a classical Be star or a high-mass X-ray binary (HMXB) system, is investigated using the first multi-epoch spectral and photometric analysis over 5.5 years (2017-2023). Optical spectroscopy from LAMOST, BeSS, and the Himalayan Chandra Telescope show large H$\alpha$ variability ($-17.6$ to $-3.2$~\AA\, in six days, returning to $-31.3$~\AA\, in 2022) and the presence of Paschen and \textsc{o}\,\textsc{i} lines, confirming active circumstellar disc dynamics. H$\alpha$ double peak profiles show weak V/R ratio variations (0.92--1.28) indicating one-armed density waves. TESS observations presents hybrid $\beta$~Cephei/slowly pulsating B-type variability with primary frequencies of 3.5~c/d and 1.5-1.6~c/d, and an outburst-like brightening with phase-dependent frequency evolution. Simultaneous anti-correlation between optical brightness and H$\alpha$ equivalent width suggest moderate disc inclination. Gaia DR3 CMD analysis rules out HD~249179 as a BeXRB but instead places it in the classical Be-star region. The BeppoSAX survey, despite previous X-ray cataloging, found no emission above the sensitivity limits. Our analysis shows that HD~249179 is most likely a classical Be star with variable circumstellar disc activity and hybrid pulsations, and that there is no strong evidence for an HMXB classification. 
}

   \authorrunning{Bhattacharyya et al. }            
   \titlerunning{Multi--epoch study of HD 249179 }  

   \maketitle

%
%
\section{Introduction}
Classical Be stars and Be/X-ray binary (BeXRBs) systems can present comparable observational characteristics. Approximately, 2/3rd of the high-mass X-ray binaries (HMXB) are BeXRBs, where the primary companion is a Be star \citep{2004MNRAS.350.1457H}. The secondary object can be a neutron star or Black hole. They usually have high orbital eccentricities. They also present strong transient X--ray outbursts (see \citealt{Reig2011}, and references therein). Classical Be stars are non--supergiant B--type stars, presents Balmer emission-line from geometrically thin circumstellar disks that rotate in a Keplerian manner \citep{2003Porter}. Disk buildup is due to mass loss caused by rapid rotation and non--radial pulsation, with possible added contribution from local magnetic activity \citep{2013A&ARv..21...69R}. The H$\alpha$ emission line is used to study disk properties. The emission strength could change over time and it provides information on formation, dissipation, and structural development of the disc \citep{2003PASP..115.1153P, 2013A&ARv..21...69R,2017AJ....153..252L}.

Be stars show H$\alpha$ variability on both short and long timescales, driven by different physical processes. Short--term variability, occurring around 0.5–2 days, is more common in early--type Be stars and is most often caused by non--radial pulsations \citep{2003A&A...411..229R}. On longer baselines, disk cycles--formation, quasi--steady emission, and dissipation--take months to years, with significant equivalent width changes usually requiring several months \citep{2011Jones}. Outbursts are much more frequent in early types \citep{2017AJ....153..252L}. In this context, \cite{2022AJ....163..226L} studied short--term TESS variability of several southern Be stars, including BeXRBs. Furthermore, in a recent study \cite{2023RNAAS...7...15H} studied the variable nature of 9 HMXB objects and showed that many of them exhibit low‑amplitude, quasi‑periodic modulations on timescales of hours to days, often with multiple frequency components.

HD 249179 (alias: AAO+28 342, 1H 0556+286) has a sparse observational history. \cite{1943Merrill} first reported hydrogen emission lines in its spectrum. \cite{1978ApJS...38..357F} catalogued the object as an X--ray source (Fourth Uhuru Catalog), where it appeared as 1H 0556+286. \cite{1997Kohoutek,1999Kohoutek} later included HD 249179 in catalogs of H$\alpha$ emission--line stars. The object was considered as possible HMXB in \cite{2000liu}. \cite{2001A&A...377..148T} conducted a BeppoSAX survey of Be/X--ray binary candidates, including HD 249179, but did not detect any X--ray emission above instrument sensitivity limits. Recent work by \cite{2022MNRAS.516.1219R} used TESS photometry to search for donor--star pulsations in HMXBs and detected quasi--periodic variability along with evidence for an outburst--like brightening event in HD 249179, though they explicitly noted uncertainty regarding its HMXB classification. Despite these observations, no dedicated spectroscopic monitoring campaign has been undertaken to understand the H$\alpha$ variability in HD 249179. Hence, the disc formation and dissipation cycles of this star, and its precise classification as a classical Be star or a Be/X‑ray binary (BeXRB), remain poorly constrained. In this paper, spectroscopic and photometric observations are combined to better understand the nature of HD 249179.

In this work, we present the first systematic analysis of rapid and long--term H$\alpha$ variability in HD 249179, based on multi--epoch LAMOST DR7 Medium--Resolution Spectroscopy (MRS) in the red arm. Over six epochs spanning from MJD 58056 (2017--10--29) to MJD 58536 (2019--02--21), we detect major EW changes from –18 \AA\, to –3.2 \AA\, over 5 days, followed by disk rebuilding in early 2019. We also complemented this data with the BeSS \citep{2011Neiner} spectra and recently followed it up utilizing HCT observation on 2022--11--16, 2022--11--27 and 2023--04--04. These observations bridge the gap between rapid variability and long--term disk--cycle studies reported in classical Be stars.

The paper is organized as follows. Section 2 explains the data sources. Section 3 presents the analysis and results. Section 4 discusses the variability properties and the implications for the circumstellar disc and system geometry. Finally, Section 5 summarizes the conclusions and outlines prospects for future multi‑wavelength monitoring.

\section{Data inventory}

\subsection{\textit{\textit{Gaia}} DR3 astrometric data}
\textit{Gaia} mission was launched by the European Space Agency (ESA) on December 19, 2013 as the successor to the Hipparcos mission. Its objective is to measure the distances, positions, space motions and perform photometry of over one billion stars in the Milky Way and beyond \citep{GaiaCollaboration2021A&A...649A...1G}. We used the newly available \textit{Gaia} DR3 photometric data and adopted the geometrical distance estimate for HD 249179 from \cite{2021Bailerjonesdr3} for this study. The colour--magnitude diagram (CMD) analysis performed in this study using the \textit{\textit{Gaia}} DR3 photometric magnitudes have strengthened our work.

\subsection{Optical spectroscopy}
\subsubsection{LAMOST DR7 Medium Resolution Spectroscopic data}
The Large Sky Area Multi‑Object Fibre Spectroscopic Telescope (LAMOST) is a 4--m reflecting type Schmidt telescope, located in Xinglong Station in China and run by the Chinese Academy of Sciences. It has an effective aperture of 3.6--4.9 m and a wide Field of View (FoV) of 5{$^{\circ}$} \citep{2012Cuia}. The focal plane hosts 4000 optical fibres, each measuring 320 microns in diameter and covering 3.3$\arcsec$ in the sky. These fibers are fed into 16 low-resolution spectrographs and then registered on to a 4k{$\times$}4k CCD. The spectra obtained for each object cover a wavelength range of 3650$-$9000 \AA, with a resolving power of $\sim$1800 in \textit{g} band \citep{2013Wang}.

We obtained a total of seven LAMOST DR7 Medium–Resolution Spectroscopic (MRS) co–added data for HD 249179, with 5 of them observed between October 29 to November 11, 2017. The rest two spectra were taken on February 17 and 21, 2019, respectively. The data were retrieved from the public LAMOST archive. We used only the red--arm spectra that covers the wavelength range between 6300 to 6800 \AA. The red--arm MRS observations have a typical spectral resolution of R $\sim$ 7500, which is good enough to study the variability of the H$\alpha$ emission line. It is observed that the co--added spectrum from February 21, 2019 presents significant H$\alpha$ variability among the individual spectra obtained on the same night, likely due to systematic error. Therefore, this particular spectrum is excluded from the present study. The remaining six spectra have a signal to noise ratio (S/N) greater than 50 in the red arm, ensuring reliable measurements.

\subsubsection{Himalayan Chandra Telescope (HCT) Spectroscopy}
We obtained a total of three low--resolution optical spectra for HD 249179 using the HFOSC instrument mounted on the 2--m Himalayan Chandra Telescope (HCT)\footnote{http://www.iiap.res.in/iao/hfosc.html} located at Hanle, Ladakh, India. The star was observed on November 16 \& 27, 2022, and April 04, 2023, respectively, depending on its visibility and available observing time. The spectral coverage is from 5800 -- 8800 \AA. The 'red region' spectra are obtained with Grism 8 (5500 -- 8800 \AA) and 167l slit, providing an effective resolution of 7 \AA~at H$\alpha$. In addition, we obtained one spectrum for the star in the 'blue region' on November 27, 2022 with Grism 7 (3400 -- 7400 \AA) and 167l slit. This is done for an attempt to estimate the spectral type of our sample star, explained in Section 3.1.3.

Standard procedures were followed for data reduction. Dome flats taken with halogen lamps were used for flat‑field correction. Bias subtraction, flat field correction and spectral extraction were performed with standard IRAF tasks. FeNe and FeAr lamp spectra were taken with the object spectra for wavelength calibration. All the extracted raw spectra were wavelength calibrated and continuum normalized using IRAF tasks. Additionally, the resulting spectra are further normalized using a python routine.

\subsubsection{Spectra from BeSS database}
The Be Star Spectral (BeSS) database\footnote{$(http://basebe.obspm.fr/basebe/)$} is a large spectral catalogue of Be stars, HAeBe stars, and B[e] supergiants \citep{2011Neiner}. It assembles spectra of those stars obtained by both professional and amateur astronomers. This database is maintained at the LESIA laboratory of the Observatoire de Paris-Meudon, France. In BeSS database, we found 4 spectra for HD 249179 taken by amateur astronomers F. Teyssier (on March 25, 2012), Umberto Sollecchia (April 15, 2015) and and Buil (November 15 and 18, 2020), which is used for this study.

\subsubsection{EW measurements and errors}

For each spectrum, the H$\alpha$ equivalent width was measured on continuum-normalised data using the IRAF \texttt{splot} task. The local continuum was defined in clean regions on both sides of the line, and the integration limits were chosen to encompass the full emission profile while avoiding neighbouring features.

The uncertainty in the equivalent width was then estimated using the description provided in \cite{2006AN....327..862V} for photon--noise dominated, continuum normalised spectra. For each spectrum we determined the signal-to-noise ratio (S/N) of the continuum from line-free windows adjacent to H$\alpha$ and used the measured line integration interval $\Delta\lambda$ and equivalent width $W_\lambda$ to compute the error $\sigma(W_\lambda)$ / error. 

All EW values and the corresponding uncertainties derived are listed in Table \ref{tab:ew_measurements}.

\subsection{Time--series photometry}

\subsubsection{TESS Photometry}

To investigate the photometric variability of HD 249179, we used observations from the Transiting Exoplanet Survey Satellite (TESS; \citealt{2015JATIS...1a4003R}). The satellite launched in 2018, employs four wide--field cameras to monitor nearly the entire sky with high precision time--series photometry. Each camera covers a FOV of 24{$^{\circ}$} × 24{$^{\circ}$} (combined field of view 24{$^{\circ}$} × 96{$^{\circ}$}), and during its continuous observations of the sky, TESS divides its coverage into sectors, each lasting approximately 27.4 days. TESS provides high-precision time-series photometry in a broad optical bandpass spanning approximately 600--1000\,nm. HD 249179 was observed in sectors 43, 44, 45, 71, and 72 between November 2021 and December 2023, all obtained in 2-min cadence mode.

In this work, we used the publicly available PDCSAP light curves produced by the SPOC pipeline, which include corrections for instrumental systematics and background effects. The data were processed using the standard TESS pipeline, which includes background subtraction, aperture photometry, and correction for instrumental systematics. Only data points free from major spacecraft systematics or flagged anomalies are retained for analysis. The cleaned and detrended light curves were used to investigate the photometric variability of the system. The results of the variability analysis are presented in Section~\ref{3.2 tess}.

\begin{figure}
    \centering
    \includegraphics[width=280pt]{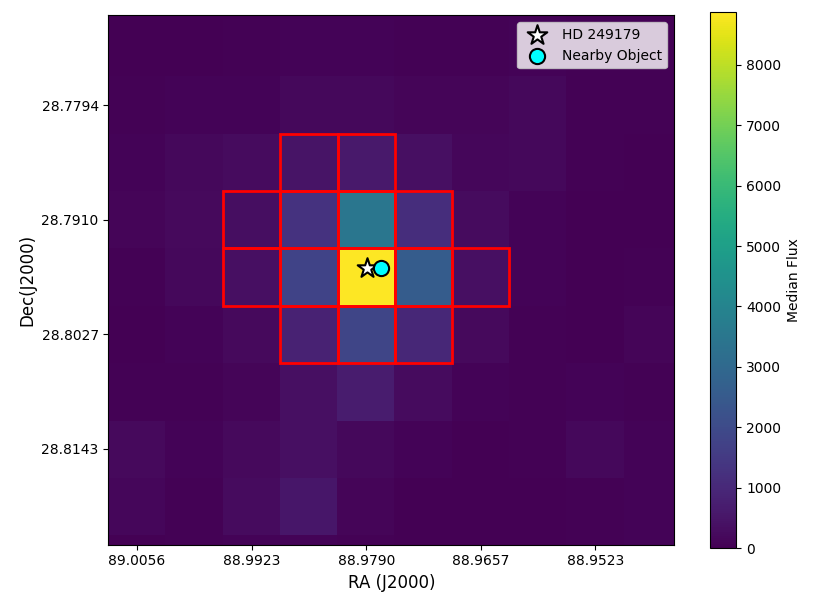}
    \caption{Median TESS sector 45 image of HD 249179 with the SPOC pipeline aperture mask overlaid (red squares). The background shows the median flux in each pixel, while the star symbol and cyan circle mark the Gaia DR3 positions of HD 249179 and the nearby object, respectively. Both sources lie inside the central TESS pixel and their fluxes are fully blended in the extracted light curves.}
    \label{fig:tess_mask_45}
\end{figure}

To visualize the possible photometric blending (see Section~3.1.1), we inspected the Sector 45 target pixel file together with the SPOC-defined optimal aperture mask (Fig.~\ref{fig:tess_mask_45}). At the TESS plate scale of approximately $21''$ per pixel, the Gaia DR3 positions of HD 249179 and the nearby source identified from the DSS2 deblending analysis both lie within the same central pixel. This represents that TESS cannot spatially resolve the pair and that the measured light curves represent the combined flux of both sources. The aperture configuration in Sector 45 is representative of the overall TESS sampling geometry for HD 249179.

\subsubsection{AAVSO and ASAS--SN light curves}
To study the long--term photometric variation in HD 249179, we have utilized two databases containing long--term photometric data for a good number of stars, namely the American Association of Variable Star Observers (AAVSO) and ASAS--SN Catalog of Variable Stars III. We collected the available V and I band light curves from AAVSO and also adopted available V-band light curve from ASAS--SN for further analysis \citep{2014Shapee, 2019Jayasinghe}.

\section{Results and Analysis}

\subsection{Characterization of the sample star}

\subsubsection{Estimating the contribution of HD 249179 on the nearby object\label{sect: deblend}}

The DSS2 red image of HD 249179 presents a nearby projected source located within approximately 5" of the target. Such a separation is comparable to or smaller than the effective apertures used in several datasets considered in this work, particularly the TESS pixels ($\approx$21"/pixel) and the LAMOST fibre diameter (3.3"). Consequently, contamination from the nearby source must be considered while interpreting both the photometric and spectroscopic measurements.

To quantify the relative flux contribution, we performed a constrained two-dimensional Gaussian deblending analysis on the DSS2 red image (see Appendix \ref{app: 1} for details). The positional coordinates of the two sources were fixed using Gaia DR3 astrometry, while the Gaussian amplitudes and widths were allowed to vary during the fitting process. The resulting model suggests that HD 249179 contributes approximately 56.1$\pm$2.5\% of the blended optical flux in the DSS2 red band, while the nearby projected object contributes the remaining $\approx$44\%.

This result indicates that the observed flux in low-spatial-resolution datasets is significantly blended. In particular, the TESS photometry represents the combined light of both objects, implying that the measured variability amplitudes are likely diluted relative to the intrinsic variability of HD 249179. Similarly, the equivalent width measurements derived from fibre-fed spectroscopy may be affected by continuum contamination from the neighbouring source. Therefore, all photometric and spectroscopic interpretations presented in this work should be considered in the context of possible blending effects.

\subsubsection{HD 249179 in the \textit{Gaia} color-magnitude diagram (CMD)}

\begin{figure}
    \centering
    \includegraphics[width=280pt]{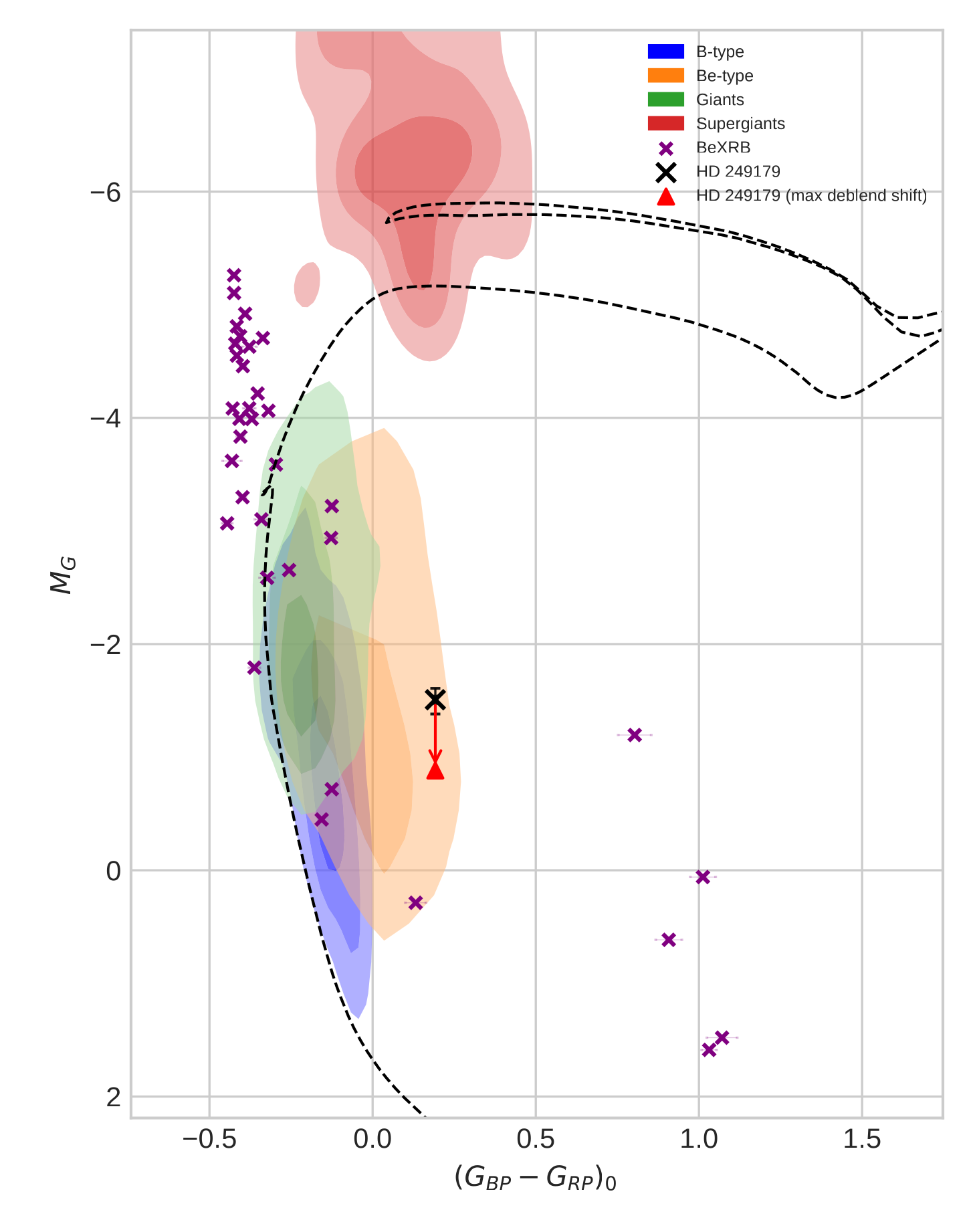}
    \caption{The CMD of HD 249179 (Black cross) having absolute \textit{Gaia} G and color corrected (BP-RP) magnitudes available from \protect\cite{GaiaCollaboration2021A&A...649A...1G}. The red arrow and filled red triangle indicates the position of HD 249179 compensating for the Gaussian deblend. The probability distribution (Gaussian fitted at three contour levels) of the B, Be stars, Giants and supergiants are shown in blue, orange, green and red shaded colors, respectively. The purple cross indicates the known BeXRB systems. The black dashed line in the plot represents the isochrone of 60 Myr with V/Vcrit = 0.4 and [Fe/H] = 0. The top black dashed line indicates the blue loop part of the same isochrone.}
    \label{fig:cmd}
\end{figure}

To investigate the evolutionary status of HD 249179 we constructed a Gaia DR3 colour–magnitude diagram (CMD) using the G, \(G_{\rm BP}\) and \(G_{\rm RP}\) photometry together with the geometric distance estimates of \cite{2021Bailerjonesdr3}. For HD 249179 we adopt the Gaia DR3 geometric distance \(d_{\rm rpgeo} = 1.60\) kpc, with reported 16th and 84th percentile confidence limits of 1.53 and 1.69 kpc, respectively. For extinction we retain the line‑of‑sight value \(A_V = 0.63\) mag from the \cite{2019Green} three‑dimensional dust map. Using these values, we compute the extinction‑corrected absolute magnitude and colour of HD 249179, which place the star at \(M_G \simeq -1.5\) and \((G_{\rm BP}-G_{\rm RP})_0 \simeq 0.19\) in the CMD (Fig \ref{fig:cmd}).

Uncertainties on the CMD position of HD 249179 are quantified explicitly in Fig \ref{fig:cmd}. The vertical error bar combines the formal Gaia G‑band photometric error with the asymmetric distance uncertainty from the Bailer‑Jones posterior. The distance interval 1.53–1.69 kpc corresponds to a range in distance modulus of \(\Delta{\rm DM} \simeq -0.10\) to \(+0.13\) mag around the central value, which dominates over the sub‑millimagnitude Gaia photometric uncertainty in \(M_G\). For the colour axis, the uncertainty is obtained from the propagated BP and RP photometric errors as
\begin{equation}
\sigma_{(G_{\rm BP}-G_{\rm RP})} =
\sqrt{\sigma_{G_{\rm BP}}^2 + \sigma_{G_{\rm RP}}^2}.
\end{equation}
Since the Gaia BP and RP errors are very small for HD 249179, the resulting uncertainty in $(G_{\rm BP}-G_{\rm RP})_0$ is also very small. As a result, the horizontal shift is not visually prominent in the CMD.

To understand the evolutionary phase of HD 249179, we added the probability distribution (Gaussian fitted at three contour levels) of previously studied B-type stars \citep{2010Huang}, Be stars (\cite{2021Bhattacharyya}, and references therein), Giant stars \citep{Hohle2010AN....331..349H} and supergiants \citep{2021Georgy}, similar to \cite{2022Bhattacharyya, 2024BSRSL..93..636B}. We also include a comparison set of well-studied BeXRB systems compiled from the literature \citep[e.g.][]{2006liu, 2019Kretschmar, 2023Neumann}, for which absolute magnitudes and dereddened colours were computed using Gaia DR3 distances, extinction corrections and the same methodology as for HD 249179. In the figure \ref{fig:cmd}, HD 249179 (black cross) lies within the broad Be‑star locus and overlaps the region where normal, non‑supergiant B‑type emission‑line star distribution. Most BeXRB systems in the comparison sample occupy higher luminosities and/or bluer colours than HD 249179, though a minority overlaps the Be‑star and pre main-sequence region, illustrating that the CMD alone cannot provide a definitive classification.

Considering the Gaussian deblending (see section \ref{sect: deblend}), HD 249179 contributes about 56 percent of the blended flux, with the remaining \(\sim 44\) percent coming from the neighbouring star. To illustrate the maximum plausible impact of any residual contamination in the Gaia solution, we mark with a red triangle the position HD 249179 would occupy if all of the blended flux were erroneously assigned to it and the true single‑star magnitude were therefore fainter by 

\begin{equation}
\Delta M_G \simeq -2.5\log_{10}(f_{\rm primary}),
\end{equation}

where $f_{\rm primary} \simeq 0.56$ is the flux fraction contributed by HD 249179. This gives $\Delta M_G \simeq 0.63$ mag. Even under this conservative assumption, and when combined with the formal distance uncertainty, the star remains inside the classical Be‑star locus and does not migrate into the region where most BeXRB systems lie. We therefore suggest that, while the CMD analysis is subject to both distance and contamination uncertainties, the Gaia DR3 position of HD 249179 is consistent with a non‑supergiant Be star and does not favour an HMXB classification based on optical colour alone.

\subsubsection{Spectral Type estimation for HD 249179}
\begin{figure}
    \centering
    \includegraphics[width=280pt]{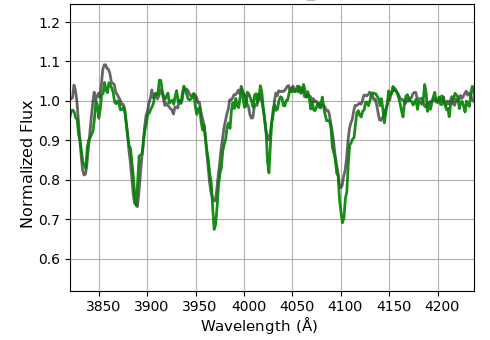}
    \caption{Fitting of the observed HCT Grism 7 spectrum (marked with green line) for HD 249179 and MILES template (marked with grey line) spectra, in the wavelength range of 3850--4200 \AA. The fit best matches with that of the B5 spectral type based on RMSE minimization using the higher--order Balmer lines, namely (H$\delta$, H$\epsilon$ and H$\zeta$)}
    \label{fig:MILES}
\end{figure}

For further characterization of HD 249179, we re‑estimated its spectral type by fitting its optical spectrum (see Figure. \ref{fig:MILES}) with templates from the MILES stellar library  \citep{2010MNRAS.404.1639V, 2006MNRAS.371..703S}. The HCT Grism 7 spectrum is used for this estimation since it covers the wavelength range between 3400 -- 7400 \AA. In order to estimate the spectral type properly, the fit is performed on higher--order Balmer lines (H$\delta$, H$\epsilon$, and H$\zeta$) as the lower--order Balmer lines can get influenced by the emission mechanism.

The fitting is done using the Root Mean Square Estimation (RMSE) minimization method, which points out the spectral type for HD 249179 to be B5. This is in agreement with the spectral type classification reported by \cite{2022MNRAS.516.1219R} and \cite{2009Belczynski}. However, we do agree that our spectral fitting using low--resolution spectra can introduce some systematic uncertainties in the fitting, with possible deviations of up to two sub--spectral classes, as highlighted by \cite{2021Anusha}.

\subsection{Photometric study of HD 249179 using TESS data \label{3.2 tess}}
Next, we studied HD 249179 using the available TESS data to better understand the photometric variability displayed by the star. The object was monitored in sectors 43, 44, and 45 consecutively during 2021, and in sectors 71 and 72 during 2023. Having a cadence of 2 minutes these light curves were acquired using the PDCSAP\_FLUX data, which represent the Simple Aperture Photometry (SAP FLUX) values after removing the systematic trends.

\begin{figure}
    \centering
    \includegraphics[width=280pt,height=\textheight,keepaspectratio]{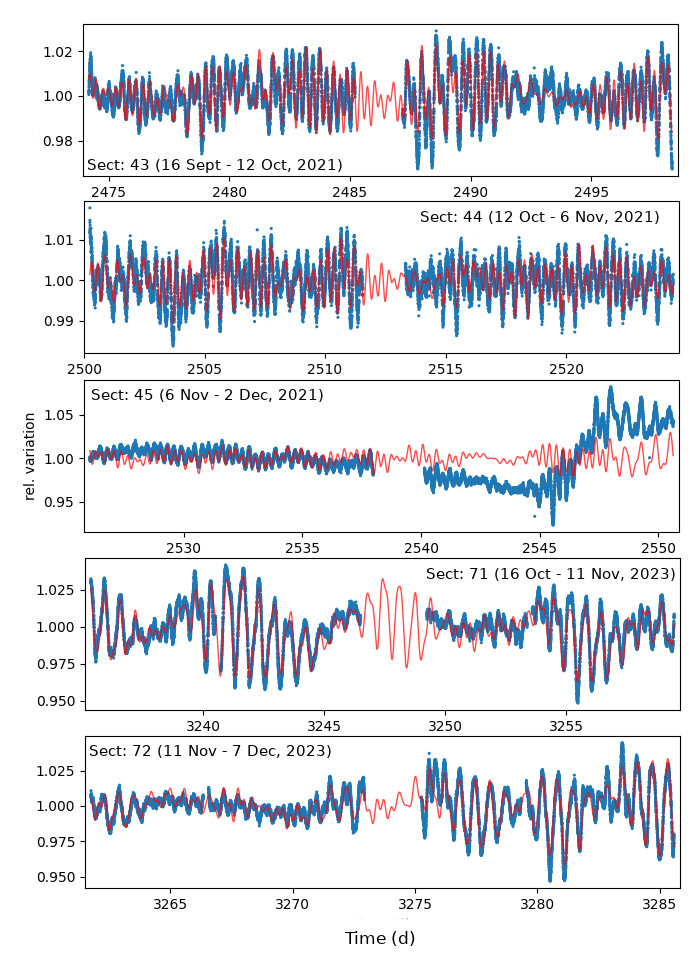}
    \caption{Normalized TESS light curves for HD 249179 spanning five observing sectors between September 2021 and December 2023. Each panel shows 2--minute cadence photometry. The data are color--coded (blue for the primary dataset and red for sine function overlay from the periodogram) to highlight different components. Sectors 43 and 44 (2021) display quasi--regular brightness variations with small amplitude (roughly 4\%). Sector 45 shows the onset of a interesting outburst event around day 2545, where the normalized flux rises from approximately 0.96 to over 1.08 within a few days, followed by a gradual relaxation. Sectors 71 and 72 (2023), obtained two years after the initial observations, demonstrate renewed variability with comparable amplitude to the early 2021 data.}
    \label{fig:TESS_LC}
\end{figure}

Figure \ref{fig:TESS_LC} shows the normalised TESS light curves for all five sectors, with the data colour‑coded to distinguish between observed points and sinusoidal fits derived from the periodogram analysis. Sectors 43 and 44 are found to display quasi--periodic modulations with amplitudes of around 4\% relative to the median flux. It is seen that Sector 45 shows interesting variability exhibiting a strong brightening event starting around day 2545 that increased the relative flux from $\approx0.95$ to $\approx1.08$ over 4 days. Now, it is observed that this rise is followed by a slow stable phase, with the flux settled to a level roughly 7\% above the precursor baseline average. Then, Sectors 71 and 72, observed two years later, show continuous variability with amplitudes reaching 3--6\% of the normalized value at 1 and displaying short--term fluctuations superimposed on the longer timescale variations.

To study the periodic properties of the light curves, we utilized the Lomb--Scargle periodogram analysis for all sectors separately with the Pyriod package based on period04 program \citep{2005CoAst.146...53L}. We iteratively used the prewhitening method to recover multiple periodic signals from the data and used a significance threshold limit 5 times of the average noise level at a window size of 35 cycles per day (c/d) to select peaks that correspond to true astrophysical signals above the noise.

\begin{figure}
    \centering
\includegraphics[width=280pt,height=\textheight,keepaspectratio]{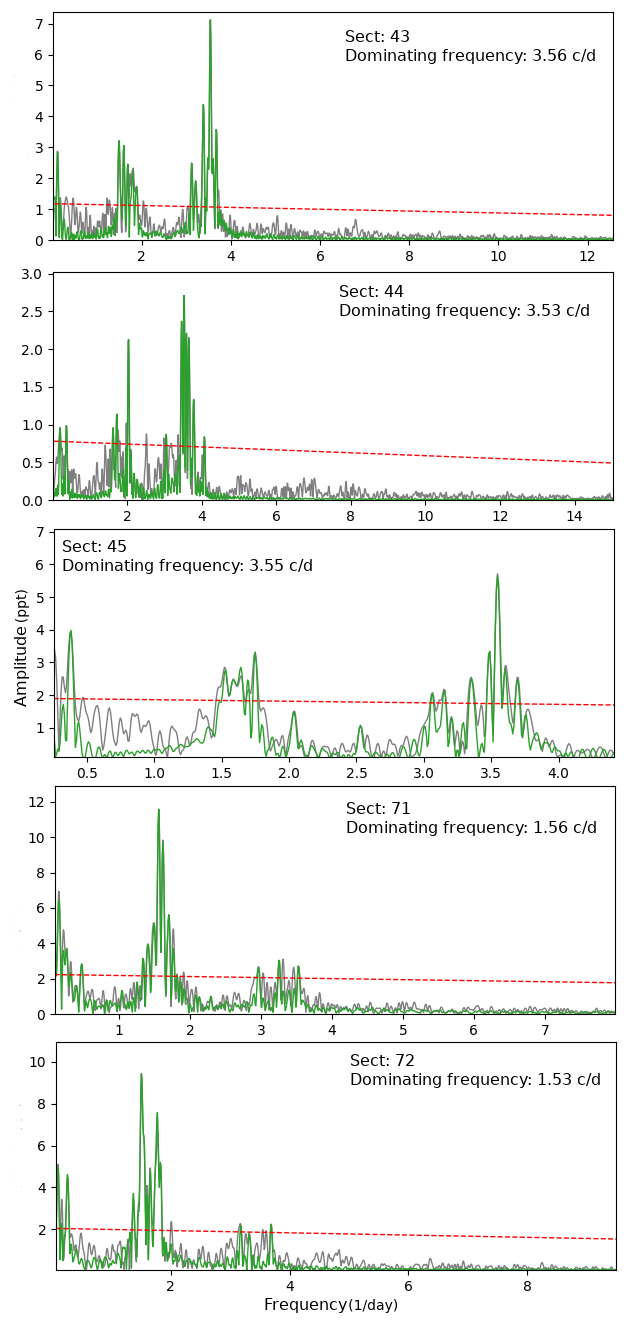}
    \caption{Lomb--Scargle periodograms of HD 249179 obtained from individual TESS sectors sectors 43, 44, 45, 71, and 72, which is covering observations from September 2021 to December 2023. The dominant frequencies detected in each sector are labeled. A significance threshold of five times the local noise level (indicated by the red dashed line) is applied to distinguish genuine periodic signals. It is visible from the periodograms that Sectors 43–45 from 2021 show prominent peaks near 3.5 c/d, while during 2023, Sectors 71–72 present lower frequencies of approximately 1.5 c/d.}
    \label{fig:Lomb}
\end{figure}

Figure \ref{fig:Lomb} presents the Lomb--Scargle periodograms for HD 249179 as obtained from individual TESS sectors sectors 43, 44, 45, 71, and 72, respectively. The periodograms show complex multi--periodic variability across all sectors. Interestingly, two prominent frequency groups are detected from our analysis. The first group consists of frequencies clustered around 3.5 c/d, which corresponds to periods near to 0.28 days. This frequency range is detected prominently in sectors 43, 44, and 45, with the dominant signal in these sectors having highest amplitudes. The 2021 observations (sectors 43--45) consistently showed multiple frequencies in this range. The second frequency group lies between 1.4 and 1.8 c/d (periods ranging between 0.56 to 0.71 days, or roughly 13--17 hours). We found that such longer-period variability is present in all five sectors but become dramatically stronger in sectors 71 and 72. In sector 71, the dominant frequency calculated is 1.564 c/d, representing a substantial increase compared to the 2021 observations. Sector 72 shows an even stronger amplitude at 1.526 c/d, though this exceptionally large value may be influenced by long--term trends or instrumental effects rather than purely periodic modulation.

The change in the dominant frequency group between the 2021 and 2023 TESS observations is notable. While the 2021 data shows the strongest power around 3.5 c/d, the 2023 observations are dominated by frequencies near 1.5--1.6 c/d. This change occurred over the two year gap between observations and indicates that the photometric behavior of HD 249179 is not constant over multi--year timescales. It is to be noted that both such frequency ranges have been commonly observed in Be stars and HMXBs with Be--type optical counterparts. The faster variations around 3.5 c/d are consistent with non--radial pulsations frequently detected in early--type Be stars, while the longer period near 0.6 days could be related to rotational modulation.

\begin{figure}
    \centering
    \includegraphics[width=280pt,height=\textheight,keepaspectratio]{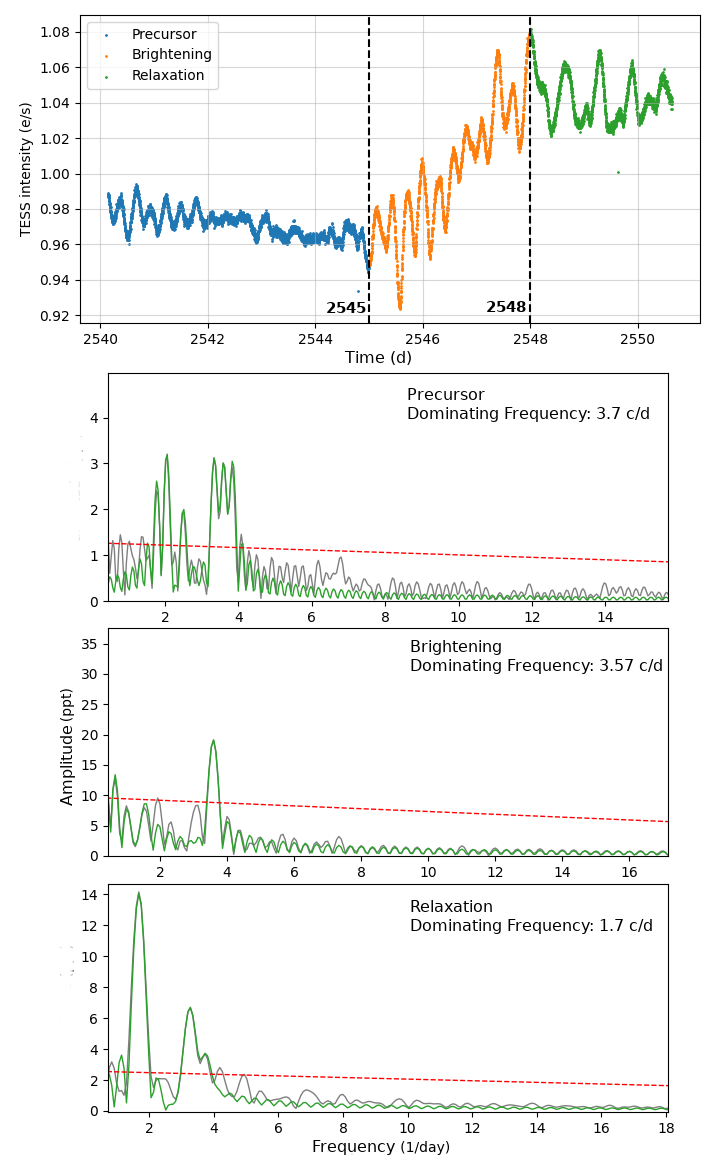}
    \caption{Analysis of the Multi--phase event in sector 45. The top panel shows the light curve segment covering days 2540–2551 and three defined phases: precursor (blue, prior to day 2545), brightening phase (orange, days 2545–2548), and relaxation phase (green, after day 2548). The three bottom panels report the corresponding Lomb--Scargle periodograms obtained from each phase. The precursor phase periodogram is having maximum amplitude at frequency of 3.7 c/d prior to the outburst. During the brightening process, the main frequency was 3.57 c/d with comparatively lower amplitude than the precursor phase. The relaxation periodogram presented the most significant frequency shifting to 1.69 c/d (0.59 days period), indicating restructuring of the photometric modulation pattern after the outburst}
    \label{fig:multiphase}
\end{figure}

As a special case in the last TESS observation of sector 45, a brightening event is observed, which was claimed to be an outburst event by \cite{2022MNRAS.516.1219R}. In Figure \ref{fig:multiphase}, we divide the outburst region of the light curve (sector 45) into three separate phases, namely precursor, brightening, and relaxation phases. The periodogram of the precursor phase (before day 2545) shows the dominant frequency at 3.7 c/d. This matches the previously observed dominating frequency. Additionally, another strong frequency at 1.9 c/d is present. During the brightening phase (days 2545--2548), the dominant frequency shifted slightly to 3.57 c/d only with significantly enhanced amplitude and the 1.9 c/d frequency become comparatively weaker. The relaxation phase (after day 2548) exhibited a strong dominant frequency of 1.7 c/d and distinguishable peak at 3.1 c/d, suggesting that the outburst event might have altered the stellar pulsation modes, possibly through disc structure or mass redistribution.

\begin{figure*}
    \centering
\includegraphics[width=\linewidth,height=\textheight,keepaspectratio]{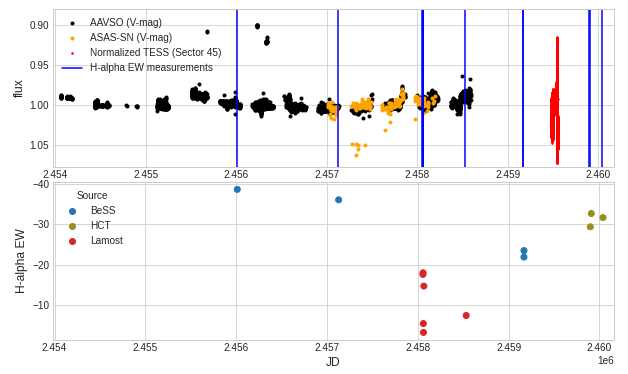}
    \caption{Variation of $H\alpha$ EW and V--mag shown for HD 249179. In the top panel, the black and orange filled circles show V--mag values obtained from ASAS and AAVSO, respectively. The vertical blue lines represent the time of spectral observations. The TESS light curve shown in red filled circles. In the bottom panel, the blue, dark yellow, and red filled circles represent the H$\alpha$ EW from spectra sources.}
    \label{fig:EW_LC}
\end{figure*}

\subsubsection{Long--term photometric behavior of HD 249179 observed from AAVSO and ASAS--SN data\label{3.3.2}}
We now further check the available light curves for this star from AAVSO to better understand its long--term photometric variations. The AAVSO light curve provides insights into the star’s photometric behavior over an extended period. In Figure \ref{fig:EW_LC} (top panel), we present the AAVSO light curve in the V and I bands, normalized using the Lightkurve package. The observations span over a period from 2010 to 2019. However, we could not perform periodogram analysis using AAVSO data due to inconsistent and sparse sampling. Nevertheless, visual inspection suggests a generally stable brightness level with no significant long--term brightening or dimming trends. The standard deviation of the AAVSO light curve in both the V and I bands is approximately 0.009 in normalized flux units. Overplotted for comparison, the ASAS--SN V-band light curve reveals a very similar photometric trend to the AAVSO data. For additional context on flux variations, the TESS sector 45 outburst phase is also shown. The bottom panel of Figure \ref{fig:EW_LC} displays the H$\alpha$ equivalent width (EW) variations measured over the same observation period, providing a direct comparison between photometric behavior and H$\alpha$ emission strength for HD 249179.

\begin{figure}
    \centering
\includegraphics[width=280pt,height=\textheight,keepaspectratio]{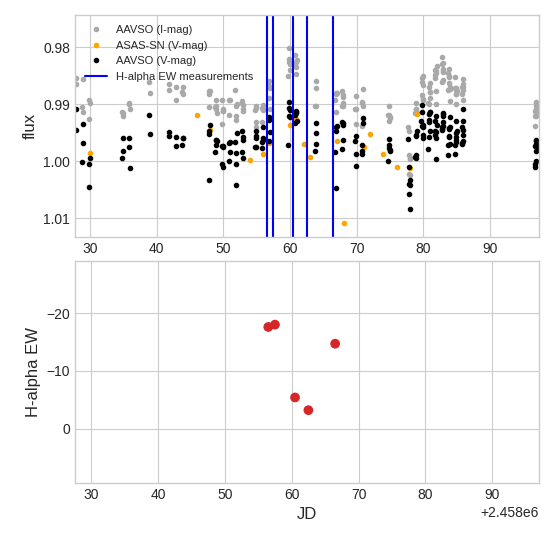}
    \caption{The AAVSO light curve variation along with the variation in the LAMOST DR7 H$\alpha$ EW strength.}
    \label{fig:anti_corr_EW_LC_lamost}
\end{figure}

In Figure \ref{fig:anti_corr_EW_LC_lamost}, a magnified view of figure \ref{fig:EW_LC} is presented, corresponding specifically to the time of the LAMOST DR7 observations in 2017. The detailed view now reveals a clear anti--correlation pattern between the photometric variations and the H$\alpha$ EW measurements. This anti--correlation is evident in both AAVSO V and I band light curves. We determined that the amplitude of the photometric variation is on the order of 0.01 to 0.02 mag in normalized flux (uncertainty of ~0.005 mag in AAVSO V band) of V band. We have also showed the AAVSO I band and ASAS--SN B band light curve with the uncertainity of 0.01 mag. Although they are showing similar trend of AAVSO V band, whereas the corresponding change in the H$\alpha$ EW, as measured from LAMOST DR7 data, is approximately 14.8 \AA. The implications of this results are discussed further in the next section.

\subsection{Spectral feature analysis for HD 249179}

\subsubsection{Study of the LAMOST DR7 and BeSS spectra}
Looking into the LAMOST DR7 data, we found that a total of 6 spectra for HD 249179 are available within a period of 16 months, i.e. between October 29, 2017 and February 17, 2019. H$\alpha$ is present in weak double--peaked emission (\textit{dpe}) in all the occasions. The presence of a circumstellar disc around this star is evident from the presence of H$\alpha$ emission all through these observations. Moreover, we detected that it has shown considerable H$\alpha$ variability over this period of 16 months. The measured H$\alpha$ EW values from different observation facilities are shown in Table \ref{tab:ew_measurements}.

It is seen from Table \ref{tab:ew_measurements} that the measured H$\alpha$ EW decreased from $-17.6 \pm 0.3$ \AA~on October 29, 2017 to a minimum of $-3.2 \pm 0.2$ \AA~on November 04, 2017 followed by a rapid increase to $-14.7 \pm 0.8$ \AA~on November 08, 2017. This means that HD 249179 has shown over 100\% variability in its H$\alpha$ EW twice within only 10 days, which might be a strong indicator of some activity, such as outburst phenomenon followed by a disc formation phase, happening in its surrounding disc. Then, the EW was found to have decreased to $-7.4 \pm 0.2$ \AA~on February 17, 2019, the last available spectrum in LAMOST DR7. This is again a sufficient decrement (around 50\%) in the H$\alpha$ EW within 15 months that might hint another outburst phenomenon or a possible disc dissipation scenario during that period.

We then checked the available 4 spectra from the BeSS. Although H$\alpha$ appears in emission in all 4 occasions, we found two out of four are high resolution (R = 12000), normalized spectra taken by Buil (using the NEWTON250 eShel2 ASI6200 setup) in 2020. So we considered these two spectra for our present study. Although, for comparison purpose we have added the EW measured from the remaining two low resolution spectra in figure \ref{fig:EW_LC}. H$\alpha$ exhibits weak \textit{dpe} profile in both these cases. The measured EW is detected to be $-21.9 \pm 1.5$ and $-23.5 \pm 1.6$ \AA~on November 15 and 18, 2020, respectively. This suggests that the star might have passed through a sufficient disc formation phase within a period of 21 months, i.e. between February, 2019 (when last LAMOST DR7 spectrum was taken) and November, 2020. Moreover, both H$\beta$ and H$\gamma$ lines were also noticed in \textit{dpe in absorption} profile on both the occasions. Apart from Balmer lines, no other prominent feature is detected.

\begin{table}
\centering
\caption{Equivalent width (EW) measurements and signal-to-noise ratio (S/R) of the H$\alpha$ line for HD 249179 from various observation facilities.}
\begin{tabular}{l l l c}
\hline
Date & Facility & S/R & EW $\pm$ error (\AA) \\
\hline
2017-10-29 & LAMOST     & 161 & $-17.6 \pm 0.3$ \\
2017-10-30 & LAMOST     & 117 & $-18.0 \pm 0.5$ \\
2017-11-02 & LAMOST     & 118 & $-5.4 \pm 0.2$ \\
2017-11-04 & LAMOST     & 108 & $-3.2 \pm 0.2$ \\
2017-11-08 & LAMOST     & 57 & $-14.7 \pm 0.8$ \\
2019-02-17 & LAMOST     & 161 & $-7.4 \pm 0.2$ \\
2020-11-15 & BeSS       & 41 & $-21.9 \pm 1.5$ \\
2020-11-18 & BeSS       & 43 & $-23.5 \pm 1.6$ \\
2022-11-16 & HCT        & 116 & $-29.4 \pm 0.8$ \\
2022-11-27 & HCT        & 176 & $-32.7 \pm 0.6$ \\
2023-04-04 & HCT        & 163 & $-31.7 \pm 0.7$ \\
\hline
\end{tabular}
\label{tab:ew_measurements}
\end{table}

\subsubsection{Study of the HCT spectra}
\begin{figure*}
    \centering
\includegraphics[width=\linewidth,height=\textheight,keepaspectratio]{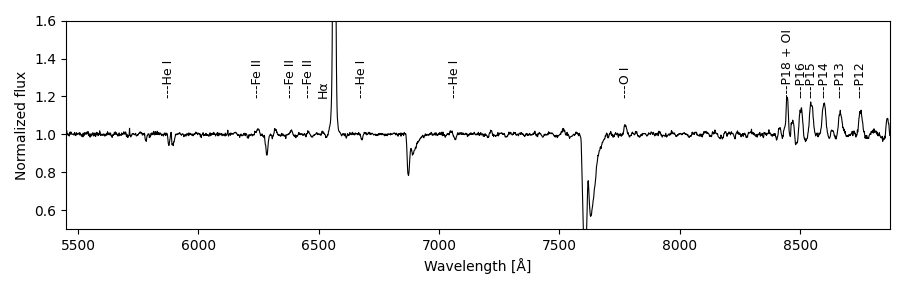}
    \caption{Representative spectrum for HD 249179 obtained by us using HCT facility with prominent emission features being marked.}
    \label{fig:representative}
\end{figure*}

A representative spectrum for HD 249179 obtained by HCT is shown in Figure \ref{fig:representative}. Looking into Table \ref{tab:ew_measurements}, it is seen that the star has shown a considerable rise in measured H$\alpha$ EW on all three dates of our observations using the HCT facility. Each spectrum has a high S/N $\geq$ 120, ensuring reliable measurements of the H$\alpha$ EW. The EW is estimated to be $-29.4 \pm 0.8$, $-32.7 \pm 0.6$ and $-31.7 \pm 0.7$ \AA~on November 16, 27, 2022 and April 04, 2023, respectively. Although a minor variability is exhibited during these 3 dates, the variation remains well within 10\%, which is within the estimated measurement uncertainty. Also, this might be one of the state with large H$\alpha$ EW for this star. We calculated the average H$\alpha$ EW from the HCT data to be -31.3 \AA, indicating the system’s H$\alpha$ emission strength has increased by at least 37\% compared to the BeSS observations during 2020, which itself is already a 3 times increment from the LAMOST DR7 observation in 2019. Our results, thus suggest that this star's disc has formed considerably since February 17, 2019, i.e. within a span of 45 months till November, 2022. Then onward, HD 249179 is possibly hosting a stable, well--developed disc. However, we cannot comment about its current disc phase due to the lack of continuously monitored data.

Interestingly, we could further probe the extended optical spectral regime for HD 249179 for the first time since our spectral coverage with HCT is more than available LAMOST DR7 and BeSS spectra. Apart from H$\alpha$, we detected prominent emission lines of Paschen series (P12 -- P18 clearly visible), emission features of O{\sc i} 7772 triplet and 8446 \AA~on all three occasions. We also found a minor variation existing in the O{\sc i} 7772 \AA~EW, being -0.6 \AA~on first two dates and -0.7 \AA~on April 04, 2023. Likewise, the P14 (8598 \AA) emission line EW is estimated to be -2.1, -2.7 and -2.0 \AA~on November 16, 27, 2022 and April 04, 2023, respectively. Here, we intentionally measured the EW of these two prominent features since P14 and O{\sc i} 7772 \AA~lines do not get blended with any other feature in low--resolution spectra like ours. In addition, He{\sc i} 5876, 6678 and 7065 \AA~lines are visible in absorption on all 3 dates of observations.

\subsubsection{V/R variability observed in HD 249179\label{3.3.3:VR}}

The H$\alpha$ profiles of HD~249179 observed in LAMOST DR7 and BeSS spectra display a weak double-peaked emission (dpe) structure in which the violet (V) and red (R) peak intensities vary between epochs. To quantify this variability, we measured the V/R ratio using eight spectra selected on the basis of spectral quality from LAMOST DR7 and BeSS.

\begin{figure*}
    \centering
\includegraphics[width=\linewidth,height=\textheight,keepaspectratio]{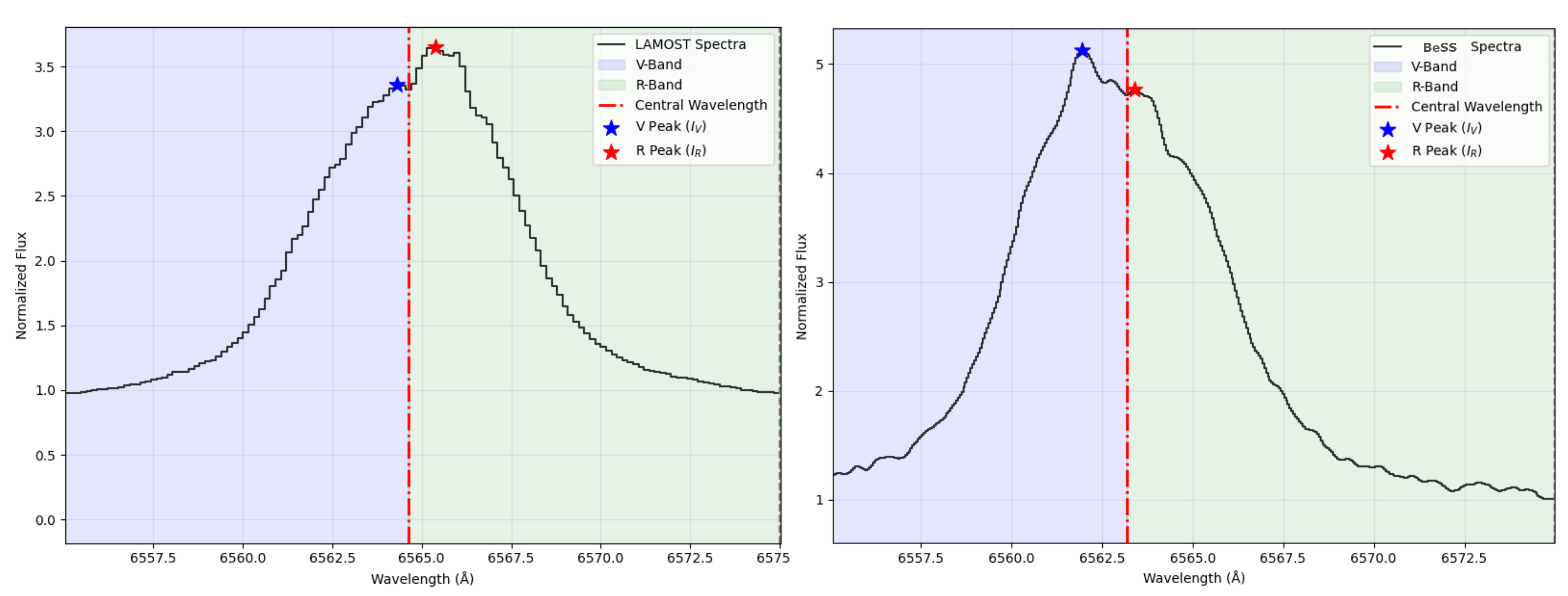}
    \caption{Representative H$\alpha$ emission profiles of HD~249179 illustrating the V/R measurement method. The blue and green shaded regions denote the violet (V) and red (R) sub-bands, respectively, separated by the central absorption minimum (red dash-dotted line). The blue and red stars mark the V and R peak intensities ($I_V$ and $I_R$) identified as the flux maximum in each sub-band. \textit{Left}: LAMOST spectrum obtained on 2017 November 8, showing V/R~$<$~1 with the red peak dominant. \textit{Right}: BeSS spectrum obtained on 2020 November 15, showing V/R~$>$~1 with the violet peak dominant. The contrast between the two panels demonstrates the sign reversal in V/R between the 2017--2019 and 2020 epochs.}
    \label{fig:vr_example}
\end{figure*}

The V and R peak intensities were measured following the empirical method described by \citet{2025A&A...697A.209C}, analogous to their Case~2 approach. The H$\alpha$ emission profile is divided into two sub-bands by the central wavelength of H$\alpha$. The maximum flux in the blue sub-band defines the V peak intensity $I_V$, and the maximum flux in the red sub-band defines the R peak intensity $I_R$. Representative examples of this measurement applied to a LAMOST spectrum (2017 November 8) and a BeSS spectrum (2020 November 15) are shown in Figure~\ref{fig:vr_example}. The left panel illustrates a case with weak V/R~$<$~1 (R~$>$~V), while the right panel shows a case with weak V/R~$>$~1 (V~$>$~R), demonstrating the sign reversal between the two epochs.

The measured peak intensities, V/R ratios, and peak separations $\Delta\lambda$ are listed in Table~\ref{tab:vr_peaksep}. Between 2017 and 2019, V/R remained below unity with values ranging from 0.92 to 0.98, indicating a consistent dominance of the red peak over six LAMOST epochs. In November 2020, the ratio increased to 1.28 and 1.08 in the two BeSS spectra, signifying that the violet peak became stronger than the red peak at this epoch. The contrast between the two regimes is visible directly in Figure~\ref{fig:vr_example}: the 2017 LAMOST profile shows a higher R peak, while the 2020 BeSS profile shows a higher V peak.

\begin{table}
\centering
\caption{V and R peak intensities, V/R ratio, and peak separation ($\Delta\lambda$) for HD~249179 at selected epochs. All fluxes are continuum-normalised.}
\begin{tabular}{l l c c c c}
\hline
Date & Instrument & V$_I$ & R$_I$ & V/R & $\Delta\lambda$ (\AA) \\
\hline
2017-10-29 & LAMOST & 3.31 & 3.56 & 0.93 & 1.2 \\
2017-10-30 & LAMOST & 3.36 & 3.64 & 0.92 & 1.1 \\
2017-11-02 & LAMOST & 1.8 & 1.91 & 0.94 & 1.66 \\
2017-11-04 & LAMOST & 1.39 & 1.46 & 0.95 & 1.8 \\
2017-11-08 & LAMOST & 3.07 & 3.27 & 0.94 & 1.7 \\
2019-02-17 & LAMOST & 2.12 & 2.16 & 0.98 & 1.4 \\
2020-11-15 & BeSS & 4.88 & 3.82 & 1.28 & 3.2 \\
2020-11-18 & BeSS & 5.12 & 4.76 & 1.08 & 1.5 \\
\hline
\end{tabular}
\label{tab:vr_peaksep}
\end{table}

Variations in the V/R ratio in classical Be stars are generally linked to one--armed density waves or global oscillations in the circumstellar disc \citep{2013A&ARv..21...69R}. The observed shift from V/R value lesser than 1 to greater than 1 suggests changes in the disc’s density distribution, possibly the evolution of asymmetric structures as the disc formation occurred following the disc dissipation phase in 2017. However, the overall V/R ratio remained close to unity, ranging only between 0.92 and 1.28, indicating the disc was mostly mildly asymmetric during these observations. No periodic V/R variability is evident in this dataset, a behavior consistent with many classical Be stars where V/R oscillations appear only over longer monitoring periods or under specific disc conditions \citep{2013A&ARv..21...69R}.

Peak separation, defined as the wavelength difference between the V and R peaks, was calculated for each spectrum. This value varied between epochs, reflecting changes in the structure or density distribution of the region emitting H$\alpha$. No further physical interpretation of these peak separations in terms of disk radius is analyzed here, since no measurement of the projected rotational velocity (\textit{v}sin\textit{i}) of HD 249179 is currently available. Instead, attention is required to the measured spectral properties, which are summarized in the table \ref{tab:vr_peaksep}.

The absence of pronounced V/R variability and the stability in peak separations suggest that HD 249179's disc remained largely uniform during these years, with only moderate changes in disk structure. This behavior reflects a common trend among Be stars in periods dominated by disk rebuilding or disk quiescence \citep{2013A&ARv..21...69R}.

\subsubsection{Impact of the nearby source on the spectra}

Considering for the fibre‑fed LAMOST spectra, both HD 249179 and the nearby source fall inside the 3.3" aperture, and the measured continuum therefore contains flux from both objects. The nearby object has not been spectroscopically classified, and its spectral type, luminosity class, and possible photometric variability remain unknown. Consequently, we cannot study its continuum contribution explicitly. Our analysis therefore assumes that the neighbouring source contributes only a continuum component over the H$\alpha$ spectral region and does not produce significant H$\alpha$ emission. Under this assumption, the measured equivalent widths represent lower limits to the intrinsic EW of HD 249179, diluted by a factor comparable to the flux fraction f$\approx$0.56 derived from the DSS2 deblending. However, no information is currently available regarding the possible variability of the nearby object, we cannot verify whether its continuum contribution remains strictly constant over the observing interval. Nevertheless, unless the companion itself presents significant photometric and spectroscopic variability, its continuum contribution is expected to remain approximately constant, affecting primarily the absolute EW values rather than the observed epoch-to-epoch relative variations. We therefore took this assumption as a potential source of systematic uncertainty, although it is unlikely to differ our primary conclusions concerning the large-amplitude H$\alpha$ variability observed in HD 249179.

For the HCT spectra, the slit width is 1.67", whereas the projected separation between HD 249179 and the nearby object is $\approx 5$". Under typical seeing conditions the companion is therefore largely excluded from the slit, and the HCT spectra are expected to suffer much less contamination than the LAMOST fibre data.

In contrast, the BeSS spectra obtained with the NEWTON 250\,mm f/4.5 telescope and the eShel spectrograph use a fibre with a diameter of about $\approx$10.3", which comfortably include both sources. In this configuration the situation is similar to the LAMOST case. The BeSS H$\alpha$ equivalent widths should thus also be regarded as lower limits to the intrinsic values for HD 249179.

\section{Discussion}

\subsection{Photometric variability characteristics}
\label{4.1 photometric tess}

The TESS photometric analysis confirms that HD~249179 presents short-timescale photometric variability. We detected two frequency groups, one having a range of around 3.5 c/d and a lower frequency range near 1.5 c/d. The 3.5 c/d frequency parameter is within the range of typical $p$--mode pulsations as found in case of ($\beta$) Cephei type variables \citep{2005ApJS..158..193S}. In case of $\beta$ Cephei variables, spectral types usually range between B0 -- B2 \citep{2005ApJS..158..193S}. However, we estimated the spectral type of HD 249179 to be B5, in agreement with previous studies \citep{2022MNRAS.516.1219R}. Interestingly, certain previous studies \cite[]{2020MNRAS.494..958N, 2005ApJ...623L.145W, 2020AJ....160...32L} have also reported that some late type Be stars do exhibit pulsation properties similar to $\beta$ Cephei stars.

The low--frequency signal (near 1.5 c/d) obtained for our sample star supports pulsation behavior observed in case of Slowly Pulsating B (SPB) type stars \citep{1991A&A...246..453W}. Such behavior is commonly observed in majority of Be stars too \citep{2002ASPC..259..196D}. The presence of additional high--frequency signals in the TESS data as detected by us do suggest that HD 249179 might be a hybrid SPB or $\beta$ Cephei type pulsator \citep{2017AJ....153..252L}.

In addition, we found that this star exhibits quasi--periodic modulations and the measured period range using all TeSS sectors is approximately 0.1 -- 1.0 days. These measurements are consistent with the study by \cite{2022MNRAS.516.1219R}, which strongly supports its possible HMXB nature. As mentioned in Section 1, this star has already been catalogued as a HMXB by different authors (e.g. \cite{2000liu, 2006liu, 2021Arnason}). Similar periodicity values are observed in many Be stars too, sharing a comparable variability pattern with HMXBs. All these pulsation properties provide further support that HD 249179 most probably is a Be star belonging to a HMXB system.

\subsection{Detection of a possible outburst event and comparison with known HMXBs and Be stars}

We also observed a notable brightening event for HD 249179 from the TESS sector 45 data \citep{2022MNRAS.516.1219R}, which led to its flux increment from approximately 0.95 to 1.08 (relative to median) over about four days (see Section \ref{3.3.2}). Interestingly, there are instances of H$\alpha$ emission outbursts in case of Be stars resulting in sudden increase of their H$\alpha$ EW. A fraction of known Be population do show outburst behavior, as highlighted by \cite{2023Froebrich}. Some recent studies such as \cite{2022AJ....163..226L, 2018Labadie-Bartz, 2017AJ....153..252L} reported that such outburst phenomenon is exhibited by 28 -- 36\% of their sample Be stars using different sample sizes. However, \cite{2018Bernhard} identified a much larger fraction (73\%) among their target Be stars to show “bursts".

A similar brightening event was reported for two well known HMXB sources, RX J0440.9+4431 in TESS Sector 19 \citep{2022A&A...667A..18R} and for V725 Tau in Sector 44 \citep{2022MNRAS.516.1219R}. \cite{2022A&A...667A..18R} divided the light curve for RX J0440.9+4431 into three phases: precursor, brightening, and relaxation. Considering RX J0440.9+4431, \cite{2022A&A...667A..18R} observed changes in frequency in different phases. In our case, the observed outburst shows clear frequency evolution, similar to RX J0440.9+4431, with a low--frequency signal around 1.9 c/d emerging during the precursor phase. However, V725 Tau did not show any frequency or period changes. Comparing with the existing literature and these two HMXB sources, we support the interpretation proposed by \cite{2022MNRAS.516.1219R} that our observed brightening may be associated with episodic mass ejection from the Be star HD 249179 into its circumstellar disc. Such mass ejection is a common phenomenon in case of many Be stars \citep{2025A&A...699A..82L} which are known to display occasional disc dissipation and formation episodes \cite[e.g.][]{2025A&A...699A..82L, 2021Cochetti, 2019Klement}. Well-studied classical Be stars such as $\omega$ CMa \citep{2018Ghoreyshi}, $\pi$ Aqr \citep{2025A&A...697A.209C} have also exhibited repeated mass-ejection episodes followed by disc growth and dissipation cycles, presenting that such behaviour occurs even in the absence of a compact companion. This indicates that HD 249179 too might have undergone a outburst episodes between the days 2545 and 2548, as detected from TESS sector 45 data. Furthermore, comparison with the previous studies provides additional evidence that HD 249179 is likely a Be star, since it has also shown a possible outburst event, similar to many known Be stars.

However, we agree that the comparison of our sample star with RX J0440.9+4431 and V725 Tau is too early as of now. Literature review points out that both of them are well-studied objects. Extensive multi-wavelength monitoring has been performed for RX J0440.9+4431, confirming it as a BeXRB system having a neutron star as its binary companion \citep{2022A&A...667A..18R, 2022A&A...665A..31F}. Therefore, classification of HD 249179 as a BeXRB only through photometric studies using TESS data is not possible. Hence, we now look into the results obtained from our multi--epoch optical spectroscopic study.

\subsection{Implications of notable H$\alpha$ and V/R variability}
\begin{figure}
    \centering
    \includegraphics[width=280pt]{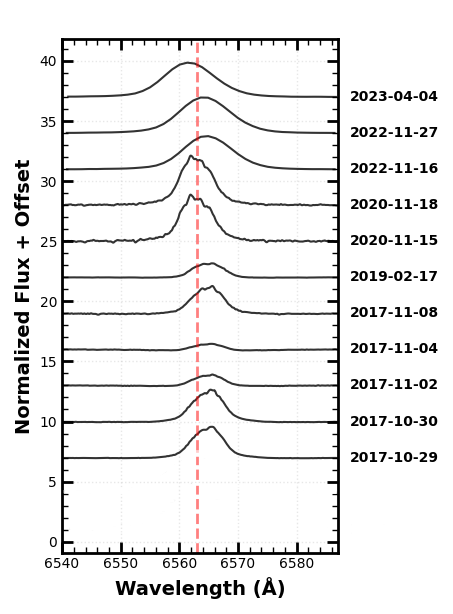}
    \caption{The evolution of the H$\alpha$ emission line profile as observed by LAMOSST DR7, Buil (BeSS) and us using HCT during a 5.5 year period, i.e. within October 29, 2017 and April 04, 2023. All spectra are normalized and vertically offset for presentation purpose. The vertical dashed red line marks the rest wavelength of H$\alpha$ at 6563 \AA\,. The observations span multiple instruments: LAMOST DR7 MRS (2017, 2019), BeSS (2020), and HCT (2022--2023). A clear weakening of the emission is evident during November 2017, followed by gradual recovery in subsequent years. The profile change in 2017 and 2019 to 2020 reflects the changing structure and density distribution of the circumstellar disk.}
    \label{fig:epoch}
\end{figure}

From our spectral feature analysis, we found that HD 249179 has undergone a notable variability in its H$\alpha$ emission strength (i.e. EW) over the period of 5.5 years, i.e. between October, 2017 and April, 2023. At first, a rapid decline of H$\alpha$ EW from $-17.6 \pm 0.3$ \AA~on October 29, 2017 to $-3.2 \pm 0.2$\AA~on November 04, 2017 (i.e. within only 6 days) was noticed while studying the LAMOST DR7 spectra. This was immediately followed by a rapid rise of H$\alpha$ EW to $-14.7 \pm 0.8$ \AA~on November 08, 2017, i.e. only within 4 days. Such a dramatic decline and rise of H$\alpha$ EW implies that the star most probably passed through an outburst event followed by a fast disc build--up phase. Then when observed with HCT during November, 2022 -- April, 2023, we determined that its average H$\alpha$ EW increased by around 37\% compared to the BeSS observations during November, 2020, which itself is already a 3 times increment from the LAMOST DR7 observations in February, 2019. This is suggestive of the fact HD 249179 most probably again passed through a considerable disc formation phase since February 17, 2019, i.e. within a span of 45 months till November, 2022. Then onward, the star has shown little variations (within 10\%) in H$\alpha$ EW, indicative of possessing a stable, well developed disc as per our observations. The evolution of the H$\alpha$ emission line for HD 249179 is shown in Figure \ref{fig:epoch}.  

Now, it is known that variability in emission line profiles is a common property of almost every Be star. Many Be stars are widely known to exhibit in their spectra either short--term variations occurring on timescales of hours to months \cite[e.g.][]{2017Paul, 2003Porter, 1996Sterken, 1982Baade} or long--term variability which occurs on timescales of years to decades \cite[e.g.][]{2002Miroshnichenko, 1994Mennickent, 1991Mennickentb}. In extreme cases, complete dissipation or formation of their circumstellar discs are also exhibited by many known Be stars, as reported by different previous literature \cite[e.g.][]{2021Marr, 2021Cochetti, 2019Klement, 2018Ghoreyshi}. Moreover, we found that the H$\alpha$ EW for HD 249179 in every occasion to be lesser than 40 \AA, as is mostly in case of Be stars irrespective of their environments \cite[]{2023Jagadeesh, 2021Banerjee}. 

Apart from the detected H$\alpha$ EW variability, we also noted weak V/R variations for this star (as described in Section \ref{3.3.3:VR}). This is a clear indication that the star is aligned with the Earth having a moderately inclined geometry, neither pole-on nor edge-on. In an important study, \cite{1991Okazaki} demonstrated that such V/R variations do occur in case of Be stars as a result of one--armed density waves or global m = 1 oscillations propagating through the circumstellar discs. Certain other studies such as those of \cite{Reig:2000} and \cite{Reig:2005qa} found correlations between V/R variations and infrared photometry in case of BeXRBs. These findings along with numerous studies from the existing literature, strongly support the classification of HD 249179 as a Be star with an active circumstellar disc.

\subsection{Correlation study from spectral and photometric analysis}

The observed anti--correlation between optical brightness (AAVSO and ASAS--SN) and H$\alpha$ equivalent width during the 2017 LAMOST observations provides important information on the disk dynamics (see figure. \ref{fig:anti_corr_EW_LC_lamost}). When circumstellar disks grow in density and size, they can show increased Balmer emission which enhances optical brightness, but disk opacity can cause absorption and reddening that suppress continuum flux \citep{2013ApJ...765...41S, 1983HvaOB...7...55H}. \cite{2013ApJ...765...41S} demonstrated that the correlation sign between H$\alpha$ emission and visual magnitude depends critically on disk inclination and scale height. Edge--on systems (shell stars) tend to show anti--correlations where increasing emission corresponds to fading as the disk obscures more stellar light. Pole--on systems typically show positive correlations where disk emission contributes to overall brightness without significant obscuration \citep{2013ApJ...765...41S, 1983HvaOB...7...55H}. Considering the LAMOST 2017 data, the anti--correlation observed in HD 249179, may indicate a higher inclination viewing geometry. As the emission strength decreases, the star becomes less obscured by its disc, leading to an overall increase in brightness.

One limitation of our present work is the unresolved nearby source, whose spectral feature remain unknown. However, its continuum contribution is expected mainly to dilute the measured EWs, future high spatial-resolution spectroscopy will be necessary to determine its nature and quantify its influence on the observed spectra.

\section{Conclusion}
The primary goal of this work was to understand the nature of the poorly studied emission--line star HD~249179, which has been variously classified as a possible classical Be star and as a candidate high‑mass X‑ray binary (HMXB). To this end, we combined multi‑epoch optical spectroscopy with space- and ground-based photometry to characterise its circumstellar environment, variability, and evolutionary status.

\begin{itemize}
\item Be star classification: Multi--epoch spectroscopy suggests the presence of persistent H$\alpha$ emission in the range from $-3.2 \pm 0.2$ to $-32.7 \pm 0.6$ \AA, indicating the presence of an active circumstellar disk throughout the available observations. Double--peaked emission profiles in H$\alpha$, H$\beta$, and H$\gamma$, along with Paschen series lines, O I 7772 \AA, and O I 8446 \AA, represent the characteristic signatures for Be star disks.

\item Disk variability: The H$\alpha$ equivalent width showed large variations, from $-17.6 \pm 0.3$ $\AA$ to $-3.2 \pm 0.2$ $\AA$ within six days, then recovered to $-14.7 \pm 0.8$ $\AA$ in four days during October--November 2017. The disk rebuilt subsequently to about -32 $\AA$ at the end of 2022. The variability of the V/R ratio between 0.92 and 1.28 points to the presence of asymmetries, which possibly indicates one--armed density waves in Be stars. Generally, this behaviour is naturally interpreted as a sequence of disc dissipation and re‑formation episodes, followed by a stable high‑emission state, fully in line with the disc dynamics observed in many classical Be stars.

\item Photometric--spectroscopic correlation: The Gaia DR3 CMD places HD~249179 within the broad Be star locus and not among the bulk of known BeXRB systems. Together with the re‑determined spectral type of B5, this supports a classification as a mid‑B classical Be star. The anti--correlation of optical brightness (AAVSO and ASAS–SN photometry) and H$\alpha$ EW, as detected during the 2017 LAMOST observations, implies a higher inclination viewing geometry. While the H$\alpha$ emission becomes weaker, the star is less obscured by its disk, thereby becoming brighter in the optical. Such behavior is indeed as theoretically expected for moderately inclined Be star systems where the continuum flux is importantly affected by the opacity of the disk.

\item Photometric variability: The multi--periodic variability photometry (TESS) reveals complex, multi--periodic variability with two main frequency groups: a high‑frequency group near 3.5 c/d (period $\approx$ 0.28 days) and a lower‑frequency group around 1.4--1.8 c/d (periods $\approx$ 0.56--0.71 days). The coexistence of these signals and their evolution between 2021 and 2023 suggest hybrid $\beta$~Cephei/SPB‑like pulsational behaviour, as reported in several other Be stars (see section \ref{4.1 photometric tess}). A brightening event, observed in TESS Sector 45, demonstrated the increase of normalized flux from 0.95 to 1.08 within four days, while the frequency evolved from 3.7 c/d to 3.57 c/d and eventually to 1.7 c/d during different phases of this phenomenon (precursor, brightening, and relaxation). Such events resemble outburst event in BeXRB star RX J0440.9+4431. However, such outburst behaviour and frequency evolution are also common among isolated Be stars, and the TESS data alone do not require the presence of a compact companion.

\item Limited HMXB signature: The BeppoSAX survey found no detectable X--ray emission above noise, which gave the evidence against an accreting compact object. There are quasi--periodic modulations with periods of 0.1 to 1.0 day, but no X--ray pulsations were detected. The binary interpretation is hence highly speculative without confirmed orbital parameters or sustained X--ray activity. 

\end{itemize}

Based on the broad spectrophotometric dataset presented in this work, HD 249179 primarily shows characteristics consistent with a classical Be star. It holds persistent disk emission, clear disk variability on short timescales, multi--periodic pulsations, and outburst phenomena. While certain observational properties, specifically the TESS brightening event and rapid disk variability, show similarity to confirmed BeXRB systems, the absence of detected X--ray emission and justified orbital parameters limits the evidence for binary interaction. Based on the available data, HD 249179 is most likely a classical Be star and possibly presenting hybrid variable behavior. Further, a possible binary companion cannot be ruled out from the data presented here, but its existence is not observationally confirmed. The determination of this question, we need further multi--wavelength and high resolution (spectroscopy) follow--up observations.

\begin{acknowledgements}
This work has made use of data from the European Space Agency (ESA) mission Gaia (https: //www.cosmos.esa.int/gaia), processed by the Gaia Data Processing and Analysis Consortium (DPAC, https://www.cosmos.esa.int/web/gaia/dpac/consortium). Funding for the DPAC has been provided by national institutions, in particular the institutions participating in the Gaia Multilateral Agreement. Also, ChatGPT (GPT 5) is used for assistance in correcting typos and grammar, though full responsibility for the manuscript’s content remains our own.
\end{acknowledgements}
\newpage
\appendix                

\section{Flux Contamination Analysis and 2D Gaussian Deblending}
\label{app: 1}

\begin{table}
    \centering
    \caption{Gaia DR3 astrometric parameters for HD 249179 and the nearby projected object (Object B). The right ascension (RA), declination (DEC), proper motions (pmra, pmdec), and renormalised unit weight error (RUWE) are taken from Gaia DR3. ‘Distance (gsp)’ denotes the geometric distance estimate from \cite{2021Bailerjonesdr3} based on Gaia parallaxes.}
    \label{tab: visual binary}
    \begin{tabular}{|c|c|c|} \hline
        Object & HD 249179 & Object B\\ \hline
        RA & 88.97934 & 88.977747\\ \hline
        DEC & 28.78510 & 28.785094\\ \hline
        pmra (mas) & 0.6335$\pm$0.034 mas/yr & 0.2327$\pm$0.06 mas/yr\\ \hline
        pmdec (mas) & -2.1888$\pm$0.021 mas/yr & -2.1782$\pm$0.04 mas/yr\\ \hline
        Distance(gsp) & 1673.64$\pm$700 pc & ---- pc\\ \hline
        RUWE & 1.5339 & 2.8947 \\ \hline
    \end{tabular}
\end{table}

\begin{figure*}
\centering
\includegraphics[width=380pt]{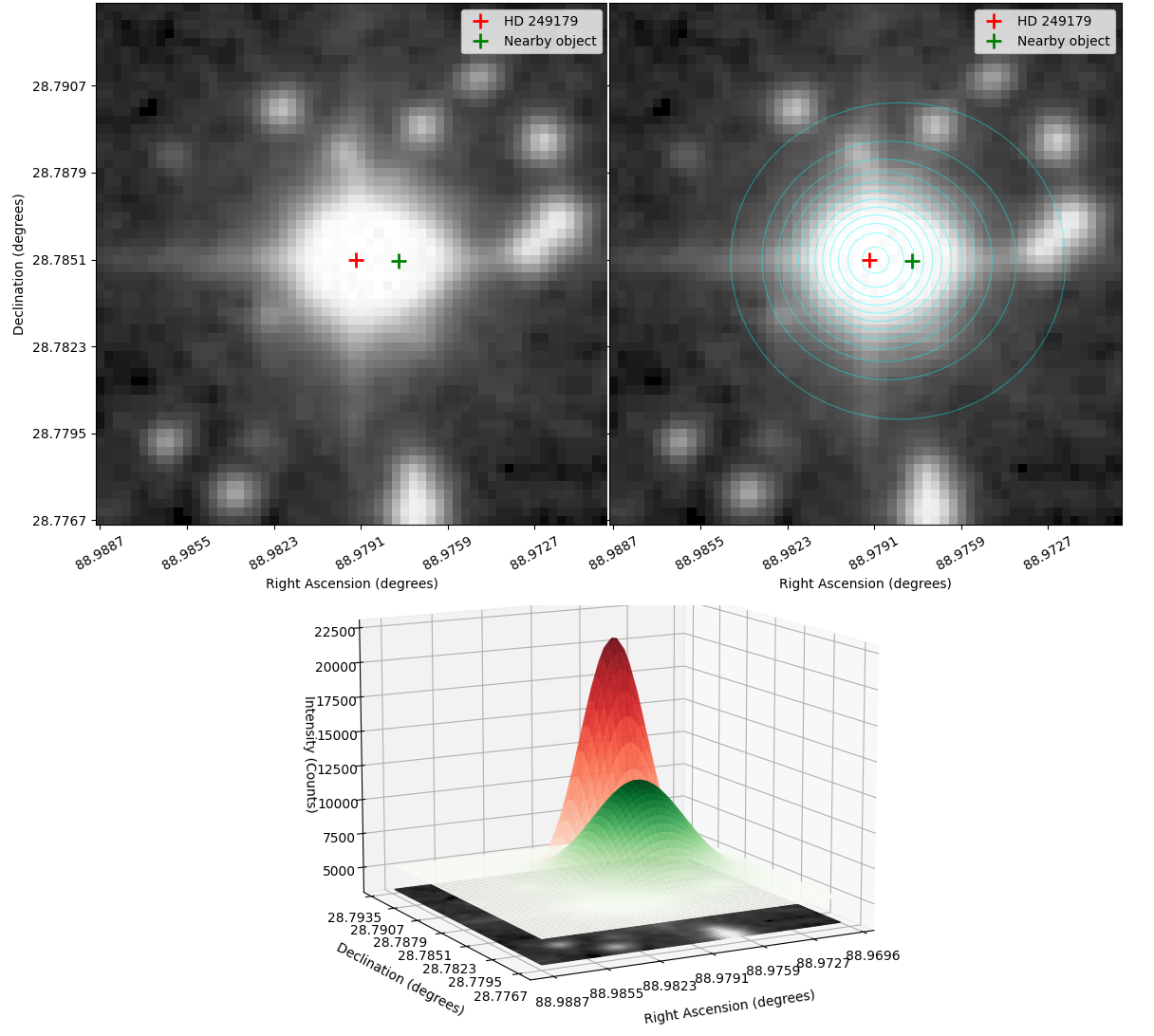}

\caption{Top: DSS2 Red image of the HD 249179 field with overlaid constrained model contours from two dimensional Gaussian fits. The top panel in the figure presents the two dimensional common distribution of two component (The cyan contour lines represent the flux from the resulting model), where red cross is the position of HD 249179 and green cross is the position of the secondary object. And the Bottom panel shows the corresponding three--dimensional plot of the deblended Gaussian models. The larger, red surface corresponds to the profile of HD 249179, while the smaller, green surface represents its fainter secondary object. The original DSS2 image has shown at the bottom.}
\label{fig:deblend}
\end{figure*}

We studied the imaging environment of HD~249179 to confirm possible observational constraints. Looking into the DSS2 image of HD 249179 (Figure \ref{fig:deblend}), it is found that two nearby objects are visible. The Figure presents the two-dimensional common distribution of two component (the cyan contour lines represent the flux from the resulting model), where red cross is the position of HD 249179 and green cross is the position of the secondary object. Gaia DR3 proper motions show significant discrepancy, particularly in the right ascension (pmra) component (0.6335$\pm$0.034 mas/yr versus 0.2327$\pm$0.06 mas/yr). Additionally, the distance estimation from \cite{2021Bailerjonesdr3} for the object B is not available. Due to these signatures and large difference in pmra, we considered that they are not associated this study. To finetune the understanding of their associationship, we need additional high cadence observations (e.g., radial velocity, spectroscopy). Their angular separation lies well within the fiber-fed spectroscopy coverage (LAMOST’s 3.3$\arcsec$ fibers) which cannot cleanly isolate the primary. Since the present study relies on the measured spectral features, especially the H$\alpha$ emission profile of this star, so it is necessary to quantify how strongly the flux from HD 249179 dominates over that of the nearby projected object considering realistic observational constraints. 

To quantify the relative flux contributions, we performed a constrained two‑dimensional Gaussian deblending of the DSS2 red image. The positional coordinates of HD~249179 and the nearby object were fixed to the values provided by astrometric measurements (Table \ref{tab: visual binary}), while the amplitudes and spread of the Gaussian components were open parameters. Figure \ref{fig:deblend} top presents the resulting two‑component model overlaid on the DSS2 image, and Figure\ref{fig:deblend} bottom shows the corresponding three -- dimensional representation of the deblended Gaussians. The larger red surface corresponds to HD~249179, and the smaller green surface corresponds to the fainter nearby object.

Using this technique, we separated the total flux into two components, directly accounting their spatial overlap. The contamination fraction was calculated by adding the model flux from the projected object falling inside a circular aperture centered at HD 249179. Importantly, error propagation is performed using Monte Carlo sampling of the fit uncertainties, producing a statistical range of estimation of the contribution. We used the following relation to calculate the contribution:

\begin{equation}
\text{Contribution \%} = \left( \frac{f_{\text{HD}}}{f_{\text{HD}} + f_{\text{PO}}} \right) \times 100\%
\end{equation}

Where $f_{\rm HD}$ and $f_{\rm PO}$ are the total modelled Gaussian fluxes from HD~249179 and the projected object, respectively. Our analysis found that approximately 56.10 $\pm$ 2.51\%  of the flux measured in the projected object’s aperture is from HD 249179. Thus, even though modest differences in coordinates exist, the optical flux in this field is dominated by HD 249179 under typical seeing conditions.

Moreover, considering LAMOST’s fiber allocation strategy \citep{2012RAA....12..805C}, which suggests a minimum separation of about 6 $\arcsec$ to prevent local over density for fiber allocation, the two objects’ separation of about 5 $\arcsec$ supports significant flux overlap. Furthermore, comparing the resolving powers, the DSS2 Red image has an angular resolution of approximately 1 $\arcsec$, enabling partial separation between the stars. However, given the coarser spatial resolution of LAMOST’s 3.3 $\arcsec$ fibers, the contamination effect is expected to be even more pronounced in the fibers, which eventually contribute in the spectra. Therefore, the contamination fraction measured from the DSS2 image should be considered a conservative lower limit for LAMOST spectra influenced by HD 249179.

\bibliographystyle{mnras}
\bibliography{bibtex}

\label{lastpage}

\end{document}